\documentclass[prb,preprint,preprintnumbers,amsmath,amssymb,floatfix]{revtex4}

\usepackage{graphicx}
\usepackage{dcolumn}
\usepackage{bm}
\usepackage{color}
\usepackage{ulem}

\begin{document}

\preprint{}

\title {New approach to extract important degrees of freedom in quantum dynamics using singular value decomposition: Application to linear optical spectrum in two-dimensional Mott insulators}

\date{\today}

\author{J. Tokimoto}
\affiliation{Nagoya Institute of Technology, Gokiso-cho, Showa-ku, Nagoya 466-8555, Japan}
\author{S. Ohmura}
\affiliation{Nagoya Institute of Technology, Gokiso-cho, Showa-ku, Nagoya 466-8555, Japan}
\author{A. Takahashi}
\affiliation{Nagoya Institute of Technology, Gokiso-cho, Showa-ku, Nagoya 466-8555, Japan}
\author{K. Iwano}
\affiliation{Graduate University for Advanced Studies, Institute of Materials Structure Science,
High Energy Accelerator Research Organization (KEK), 1-1 Oho, Tsukuba 305-0801, Japan}
\author{H. Okamoto}
\affiliation{Department of Advanced Materials Science, University of Tokyo, Chiba 277-8561, Japan}

\begin{abstract}
We propose a new approach to extract the important degrees of freedom in quantum dynamics induced by an external stimulus.
We calculate the coefficient matrix numerically, where the $i-l$ element of the matrix is the coefficient of the $l$th basis state at the $i$th discretized time in the solution of the time-dependent Schr\"{o}dinger equation induced by the external stimulus. By performing a randomized singular value decomposition of the coefficient matrix, a practically exact solution is obtained using a linear combination of the important modes, where the number of modes is much smaller than the dimensions of the Hilbert space in many cases. We apply this method to analysis of the light absorption spectrum in two-dimensional (2D) Mott insulators using an effective model of the 2D Hubbard model in the strong interaction case. From the dynamics induced by an ultrashort weak light pulse, we find that the practically exact light absorption spectrum can be reproduced by as few as 1000 energy eigenstates in the $1.7 \times 10^7$-dimension Hilbert space of a 26-site cluster.
These one-photon active energy eigenstates are classified into free holon and doublon (H-D) and localized H-D states. In the free H-D states, the main effect of the spin degrees of freedom on the transfer of a holon (H) and a doublon (D) is the phase shift, and the H and the D move freely. In the localized H-D states, an H and a D are localized with relative distances of $\sqrt {5}$ or $\sqrt {13}$. The antiferromagnetic (AF)  spin orders in the localized H-D states are much stronger than those in the free H-D states, and the charge localization is of magnetic origin. There are sharp peaks caused by excitations to the localized H-D states below the broad band caused by excitations to the free H-D states in the light absorption spectrum.
\end{abstract}

\maketitle
\section{Introduction}
The transition from a two-dimensional (2D) Mott insulator phase to a metallic or high-temperature (high-$T_c$) superconducting phase is induced by chemical doping, and the physics of doping of 2D Mott insulators has been attracting considerable research attention.\cite{2DMIdoping1,2DMIdoping2,2DMIdoping3} Mott insulators are antiferromagnetic (AF) because exchange energy can be gained when their neighboring spins are oppositely aligned, and an empty site (designated a holon, H) and a doubly occupied site (designated a doublon, D) are mobile excitations that can carry a charge in Mott insulators. The separation of the spin and charge degrees of freedom (spin-charge separation) is considered to be a basic concept that underpins various properties of one-dimensional (1D) Mott insulators.\cite{SCSWF,SCSPL1,SCSPL2} In contrast, the translation motion of an H or a D in the AF background destroys the AF order, and the coupling between the spin and charge degrees of freedom is the origin of various physical properties of doped 2D Mott insulators.\cite{SCCinDoped2DMI1,SCCinDoped2DMI2,SCCinDoped2DMI3,SCCinDoped2DMI4,SCCinDoped2DMI5,SCCinDoped2DMI6,SCCinDoped2DMI7,SCCinDoped2DMI8,SCCinDoped2DMI9,SCCinDoped2DMI10,SCCinDoped2DMI11,SCCinDoped2DMI12,SCCinDoped2DMI13}

Because H and D (H-D) pairs are generated by photoexcitation, we can investigate their coupling based on the optical properties of 1D and 2D Mott insulators. Extensive theoretical studies have been conducted on these insulators using the half-filled Hubbard and extended Hubbard models, and by using the effective Hubbard models in the strong interaction case.\cite{effM1,effM2,effM3,effM4,effM5}
The main optical properties of 1D Mott insulators have been shown to originate from spin-charge separation.\cite{Abs1DMI1,Abs1DMI2,Abs1DMI3,Abs1DMI4,Abs1DMI5,Abs1DMI6,Abs1DMI7,Abs1DMI7.5,PIM1D,Abs1DMI8}
However, it has been shown that spin-charge coupling plays an important role in the dynamics of photogenerated charges in 2D Mott insulators.\cite{Abs1DMI6,Srelax1,Srelax2,Srelax3,Srelax4,Srelax5,Srelax6,Srelax7,Srelax8,Srelax9,Srelax10} Furthermore, the light absorption spectrum of a 2D Mott insulator is essentially different from that of a 1D Mott insulator.
In the 1D Hubbard model, the light absorption spectrum mainly consists of a few discrete peaks in the small-sized clusters, and the spectrum shape remains almost unchanged when the magnitude of the on-site Coulomb interaction is varied.\cite{CM} In the 2D Hubbard model, an absorption band is formed, and the spectrum shape is strongly dependent on the magnitude of the Coulomb interaction. 
Specifically, exciton-like peaks have been found near the low-energy edge of the continuum band in the light absorption spectrum of the 2D Hubbard model.\cite{2Dabs1,2Dabs2,2Dabs3,2Dabs4,2Dabs5,2Dabs6,2Dabs7,2Dabs8,TDMRG,2Dabs9,2Dabs10,IO} Based on consideration of the parameter dependences of the peak positions and magnitudes, it has been found that these peaks are of magnetic origin.\cite{TDMRG} A few of the lowest-energy eigenstates that contribute to the exciton-like peaks have been calculated using the numerical diagonalization method and the physical properties of these eigenstates have been investigated.\cite{IO} However, it has not been possible to confirm a clear signature of the binding of an H-D pair in real space. This is partly because the energy eigenstates that contribute to the continuum band cannot be calculated using this method. The characteristics of the energy eigenstates that contribute to the exciton-like peaks are clarified by comparing them with those that contribute to the continuum band.

It has been found that the exciton-like bound states of an H-D pair exist and that there are peaks caused by excitations of these states below the continuum band in the light absorption spectrum in both the 1D and 2D extended Hubbard models when the Coulomb interaction between the different sites is strong enough.\cite{effM2,1D2Dexciton1,1D2Dexciton2,1D2Dexciton3,1D2Dexciton4,1D2Dexciton5,1D2Dexciton6,1D2Dexciton7} Therefore, the H-D binding is caused by direct Coulomb interactions in these cases, and the energy eigenstates that contribute to the exciton-like peaks in the 2D Hubbard model are different from these exciton-like bound states of an H-D pair resulting from the direct Coulomb interaction.

The strong dimensionality dependence of the light absorption spectrum is closely related to the differences in spin-charge coupling between the 1D and 2D Mott insulators. Because chemical doping of 1D Mott insulator materials is difficult, photoinduced phenomena represent important stages for investigation of the coupling.\cite{PIM1D}
However, we have not succeeded to date in understanding the origins of the characteristic features of the light absorption spectrum in a 2D Mott insulator.

We therefore propose a new approach to address this problem in this paper. We can extract the important degrees of freedom for quantum dynamics induced by an external stimulus through singular value decomposition (SVD) of the coefficient matrix for solution of the time-dependent Schr\"{o}dinger equation. When we consider an ultrashort weak light pulse to be the external stimulus, we then obtain all energy eigenstates that contribute to the light absorption spectrum. Analysis of these energy eigenstates allows us to understand the characteristic features of the spectrum. All the energy eigenstates can be calculated via exact diagonalization of the Hamiltonian in very small clusters of up to approximately 10 sites in the Hubbard model. However, finite size effects are very serious in the results obtained in these small clusters. In particular, the problem of H-D binding in real space cannot be investigated from these results. The proposed method using SVD enables us to calculate all energy eigenstates that contribute to the light absorption spectrum in much larger clusters, in which the finite-size effects are not significant. This is possible because only the important energy eigenstates are extracted using the proposed method.

\section{singular value decomposition of coefficient matrix} \label{sec:SVDCM}
The solution to the time-dependent Schr\"{o}dinger equation $|\psi (t) \rangle$ is given by the linear combination of the basis states $|\varphi _l \rangle$ as
\begin{eqnarray} \label{eq:LC}
|\psi (t) \rangle= \sum_{l=1}^{N_{\rm b}} c_l(t) |\varphi _l \rangle,
\end{eqnarray}
where $N_{\rm b}$ is the dimension of the Hilbert space and the coefficients $c_l(t)$ are complex numbers. From the solution $|\psi (\Delta t j) \rangle$ for $1 \le j \le N_{\rm t}$ with the time interval $\Delta t$ and the number of time steps $N_{\rm t}$,
we obtain the $N_{\rm t} \times N_{\rm b}$ matrix $C$ with the $j-l$ element $c_l(\Delta t j)$.
We can extract the important modes for the dynamics from the SVD of the coefficient matrix $C$: $C=U \Sigma V^{\dagger}$. Here, $U$ ($V$) is an $N_{\rm t} \times N_{\rm t}$ ($N_{\rm b} \times N_{\rm b}$) unitary matrix, and $\Sigma$ is an $N_{\rm t} \times N_{\rm b}$ diagonal matrix with positive real entries, where the number of nonzero diagonal elements is less than or equal to the rank $R$ of $C$. Using this equation, $|\psi (\Delta t j) \rangle$ is then given by a superposition of the modes $|\Phi_k \rangle$ as
\begin{eqnarray} \label{eq:PEbymodes}
|\psi (\Delta t j) \rangle= \sum_{k=1}^R U_{j,k} \sigma_k |\Phi_k \rangle,
\end{eqnarray}
where $U_{j,k}$ ($V_{l,k}$) is the $j-k$ ($l-k$) element of $U$ ($V$), $\sigma_k$ is the $k-k$ element of $\Sigma$, and
\begin{eqnarray} \label{eq:modes}
|\Phi_k \rangle = \sum_{l=1}^{N_{\rm b}} V_{l,k}^* |\varphi _l \rangle.
\end{eqnarray}
The $k$-th mode $|\Phi_k \rangle$ is given by the $k$-th column of $V$, and the $k$-th row of $U$ gives the time dependence of the coefficient for $|\Phi_k \rangle$. The modes $|\Phi_k \rangle$ are normalized and orthogonal, and $|U_{j,k} \sigma_k|^2$ gives the weight of $|\Phi_k \rangle$ in $|\psi (\Delta t j) \rangle$.
The singular values $\sigma_k$ are sorted in descending order such that $\sigma_1 \ge \sigma_2 \ge \cdots \ge \sigma_R \ge 0$, and thus the modes are numbered in order of their importance. If the values of $\sigma_k$ for $k \ge K$ are much smaller than $\sigma_1$, then $|\psi (\Delta t i) \rangle$ can be approximated by superposition of the $K$ modes as:
\begin{eqnarray} \label{eq:PEbymodes}
|\psi (\Delta t j) \rangle= \sum_{k=1}^K U_{j,k} \sigma_k |\Phi_k \rangle.
\end{eqnarray}
In many cases, $K$ is much smaller than $N_{\rm b}$, and the solution $|\psi (t) \rangle$ can then be approximated using a linear combination of far fewer important modes than $N_{\rm b}$. The important degrees of freedom for the dynamics are then extracted by SVD of the coefficient matrix.

As will be described later, we consider large coefficient matrices with $N_{\rm b} = 1.7 \times 10^7$ and $N_{\rm t} = 6000$ in this paper. SVD of these large matrices cannot be performed using conventional numerical methods. We therefore adopt a numerical method called randomized SVD (RSVD).\cite{RSVD} RSVD is a type of low-rank approximation but it provides practically rigorous SVD for the modes where the singular values are not negligible.

\section{Model} \label{sec:M}
We applied the RSVD method to understanding of the characteristic features of the light absorption spectrum of 2D Mott insulators. We consider the half-filled Hubbard model on a two-dimensional square lattice, which is given by
\begin{eqnarray} 
H &=& \hat{K}+\hat{V},  \label{eq:HH} \\
\hat{K} &=& -T \sum_{<n,m>} \sum_{\sigma}
( c_{n, \sigma}^{\dag} c_{m, \sigma} + {\rm H.c.} ),  \label{eq:TT} \\
\hat{V} &=& U \sum_n c_{n, \uparrow}^{\dagger} c_{n, \uparrow} c_{n, \downarrow}^{\dagger} c_{n, \downarrow},  \label{eq:CT}
\end{eqnarray}
where $c_{n, \sigma}$ and $c_{n, \sigma}^{\dag}$ are the 
annihilation and creation operators, respectively, for an electron of
spin $\sigma$ at a site $n$, $<n,m>$ are pairs of nearest-neighbor
sites, $-T$ is the transfer integral between these nearest-neighbor sites, and
$U$ is the on-site Coulomb repulsion energy.

The copper oxides La$_2$CuO$_4$, Nd$_2$CuO$_4$, and Sr$_2$CuO$_2$Cl$_2$ are all 2D Mott insulator materials, and the experimentally obtained optical conductivity\cite{2DabsE1,2DabsE2,SCCinDoped2DMI13} is reproduced well by the time-dependent density-matrix renormalization group (tDMRG) calculation performed in a half-filled 2D Hubbard model with a system size of $N=36$.\cite{TDMRG}
We have calculated the light absorption spectrum $\alpha (\omega)$ using the exact diagonalization method in the 2D Hubbard model with system sizes of $N=10$, 16, 18, 20, 26, and 32, which are all compatible with the periodic boundary condition. For $N \le 20$, $\alpha (\omega)$ changes greatly with variations in $N$, and the $\alpha (\omega)$ differs significantly from that obtained for $N=36$ via the tDMRG calculation. When $N \ge 26$, the spectrum for $N=36$ can be reproduced reasonably, but SVD of the coefficient matrices is practically impossible to perform for $N \ge 26$, even when using the RSVD method.

We therefore adopt the effective Hamiltonian for the Hubbard model, which is valid in the strong correlation case where $U \gg T$, in this paper.
RSVD of the coefficient matrices can then be performed for $N=26$ by using the effective model.
For the special case where $T=0$, the energy eigenvalues
simply take the values of $mU$, where $m$ is the number of doubly occupied
sites, and a huge number of energy eigenstates with
different spin configurations are then degenerated at each energy level.
For a finite $T$ that satisfies the condition $U \gg T$,
the degeneracy is lifted, and each discrete energy level becomes
an energy band.
When the effects of the transfer term $\hat{K}$ to the second order in $T/U$ are taken into account,
the effective Hamiltonian $H_{\rm eff}$ for the Hubbard model is given by \cite{effM1}
\begin{eqnarray} 
H_{\rm eff} = \sum_{m} ( U P_m + H_{\rm eff}^{(m)}), \label{eq:Heff}
\end{eqnarray}
where $H_{\rm eff}^{(m)}$ is the effective Hamiltonian for the states in the ($m+1$)th lowest energy band, which have $m$ doubly occupied sites that are given by
\begin{eqnarray} 
H_{\rm eff}^{(m)} = P_m \hat{K} P_m - U^{-1} P_m \hat{K} P_{m+1} \hat{K} P_m
                   + U^{-1} P_m \hat{K} P_{m-1} \hat{K} P_m, \label{eq:Heffm}
\end{eqnarray}
and $P_m$ is the projection operator onto the Hilbert subspace
$S_m$ for states with $m$ doubly occupied sites.

In Mott insulators, the H and the D are mobile excitations that may carry a charge. The first term of $H_{\rm eff}^{(m)}$ describes the transfer of an H and a D to singly-occupied nearest-neighbor sites.
The second term describes the AF Heisenberg spin-spin
interaction that occurs between nearest-neighbor sites with the coupling constant
$J=4T^2/U$ and a virtual three-site transfer. 
The third term is inherent in photoexcited states and
describes several different transfer processes of an
H-D pair located at nearest-neighbor sites. 
This effective Hamiltonian represents an extension
of the well-known $t$-$J$ Hamiltonian to multiphoton excited states.
There are two energy scales in $H_{\rm eff}^{(m)}$. The charge transfer term $P_m \hat{K} P_m$ is proportional to $T$,
while the second and third terms are proportional to $T^2/U$.

We calculate the light absorption spectrum $\alpha (\omega)$ via the Lanczos method using $H_{\rm eff}^{(1)}$ and using the original Hubbard Hamiltonian. We find that the results agree well for $U/T \ge 10$, which indicates that $H_{\rm eff}^{(1)}$ is valid for use as an effective Hamiltonian to consider the characteristic light absorption spectrum of a 2D Mott insulator in the strong correlation case.

\begin{figure}[thbp]\centering
  \includegraphics[width=70mm]{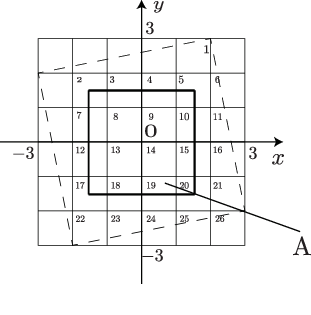}
  \caption{The shape of the $N=26$ cluster is indicated by the dashed lines. Each site is marked with its individual site number. The square with the bold sides represents the subsystem $A$.}
  \label{fig:lattice}
\end{figure}

We solve the time-dependent Schr\"{o}dinger equation, 
\begin{eqnarray} \label{eq:Schr}
i\frac{\partial}{\partial t}|\psi (t) \rangle= H_{\rm eff}^{(1)}  |\psi (t) \rangle,
\end{eqnarray}
with the initial state
\begin{eqnarray} \label{eq:Icond}
|\psi (0) \rangle = {\hat J}|\phi_0 \rangle.
\end{eqnarray}
The operator ${\hat J}=\hat{\bm{J}} \cdot \bm{e}$ is a component of the current operator $\hat{\bm{J}}$ along the direction of the unit vector $\bm{e}$, and $\hat{\bm{J}}$ is given by
\begin{eqnarray} \label{eq:J}
\hat{\bm{J}} = -iT \sum_{<n,m>, \sigma}
 (\bm{r}_m-\bm{r}_n)(c_{n, \sigma}^{\dag} c_{m, \sigma}-{\rm H.c.}),
\end{eqnarray}
where $\bm{r}_n$ is the position vector of the site $n$, as shown in Fig.~\ref{fig:lattice}, and the unit of distance is set to be the lattice spacing. The ground state $|\phi_0 \rangle$ and the energy eigenvalue $E_0$ of $|\phi_0 \rangle$ are obtained by diagonalizing $H_{\rm eff}^{(0)}$. $H_{\rm eff}^{(0)}$ is the Heisenberg spin Hamiltonian and $|\phi_0 \rangle$ has the AF spin order.
The time-dependent Schr\"{o}dinger equation is solved here by using the time-dependent exact diagonalization method.
Because the initial state $|\psi (0) \rangle$ is in the subspace $S_1$, as shown by Eq.~(\ref{eq:J}), the solution is always in $S_1$ and the time evolution is only driven by $H_{\rm eff}^{(1)}$.

When the ground state is excited by a pulse given by the vector potential $\bm{A}(t) = \bm{e} A \delta (t)$, the photoexcited state immediately after pulse irradiation is given by $A {\hat J}|\phi_0 \rangle$ to the first order in the amplitude $A$. Therefore, the solution $|\psi (t) \rangle$ considered here with the initial state given by Eq.~(\ref{eq:Icond}) gives the dynamics when the ground state $|\phi_0 \rangle$ is excited by an ultrashort weak light pulse that is polarized along the direction of $\bm{e}$.
The initial state is given by a linear combination of the energy eigenstates $|\phi _n \rangle$ ($1 \le n \le N_{\rm b}$) of $H_{\rm eff}^{(1)}$ that have the same symmetry as the initial state. We consider the translation, rotation, and charge conjugation symmetries, and then the dimension $N_{\rm b}$ of the Hilbert space is $1.7\times10^7$ when $N=26$.

The initial state is given by a linear combination of $|\phi _n \rangle$ as:
\begin{eqnarray} \label{eq:Initstate}
{\hat J}|\phi_0 \rangle = \sum_{n=1}^{N_{\rm b}} J_{n,0} |\phi _n \rangle,
\end{eqnarray}
where
\begin{eqnarray} \label{eq:TDM}
J_{n,0}=\langle \phi _n |{\hat J} |\phi _0 \rangle,
\end{eqnarray}
is the transition dipole moment when the light is polarized along $\bm{e}$.
The solution $|\psi (t) \rangle$ to the time-dependent Schr\"{o}dinger equation is therefore given by
\begin{eqnarray} \label{eq:LCEE}
|\psi (t) \rangle= \sum_n  J_{n,0} \exp(-iE_nt) |\phi _n \rangle.
\end{eqnarray}
where $E_n$ is the energy eigenvalue of $|\phi _n \rangle$.

By comparing Eq.~(\ref{eq:LCEE}) with Eq.~(\ref{eq:PEbymodes}), we find from the uniqueness of the SVD that
$|\Phi_k \rangle = |\phi _k \rangle$, $\sigma_k = {\sqrt N_{\rm t}} J_{k,0}$, and 
\begin{eqnarray} \label{eq:Uik}
U_{j,k}= \frac{1}{\sqrt N_{\rm t}} \exp(-iE_k \Delta t j),
\end{eqnarray}
hold within the limits of $\Delta t \to 0$ and $T_{\rm I} = \Delta t N_{\rm t} \to \infty$.
Here, the energy eigenvalues $E_k$ and the eigenstates $|\phi _k \rangle$ are numbered such that the values of $\sigma_k$ are sorted in descending order.

Therefore, if we use a small enough $\Delta t$ and a large enough $T_{\rm I}$, we obtain $|\phi _k \rangle$, $E_k$ and $ J_{k,0}$ from the SVD of the coefficient matrix.
The energy eigenvalue $E_k$ is obtained from the peak energy of the Fourier transform of $U_{j,k}$.
Because $U_{j,k}$ is given based on the finite time range $0 \le t \le T_{\rm I}$, the Fourier transform is performed using the Blackman window function $w(t)=0.42-0.5\cos(2\pi t /T_{\rm I})+0.08\cos(4\pi t /T_{\rm I})$, and the Fourier component with $\omega$ is then given by
\begin{eqnarray} \label{eq:UFT}
{\hat U}_{k}(\omega) = \sum_{j=1}^{N_{\rm t}} w(\Delta t j) U_{j,k} \exp(i\omega \Delta t j).
\end{eqnarray}
Here, we note that if some eigenstates have the same (or extremely close) singular values or eigenvalues, one cannot distinguish one from the others: the modes are mixed up with arbitrary coefficients of a linear combination.

If the $\sigma_k$ values are negligible for $k > K$, then $\alpha (\omega)$ is given approximately by 
\begin{eqnarray} \label{eq:Abs}
\alpha (\omega)=-{1 \over \pi}\sum_{k=1}^{K} \Im [{{|J_{k,0}|^2} \over {\omega-U-E_k+E_0+i\epsilon}}],
\end{eqnarray}
where $\epsilon$ is the broadening.

\section{Results} \label{sec:Results}
\begin{figure}[thbp]\centering
  \includegraphics[width=120mm]{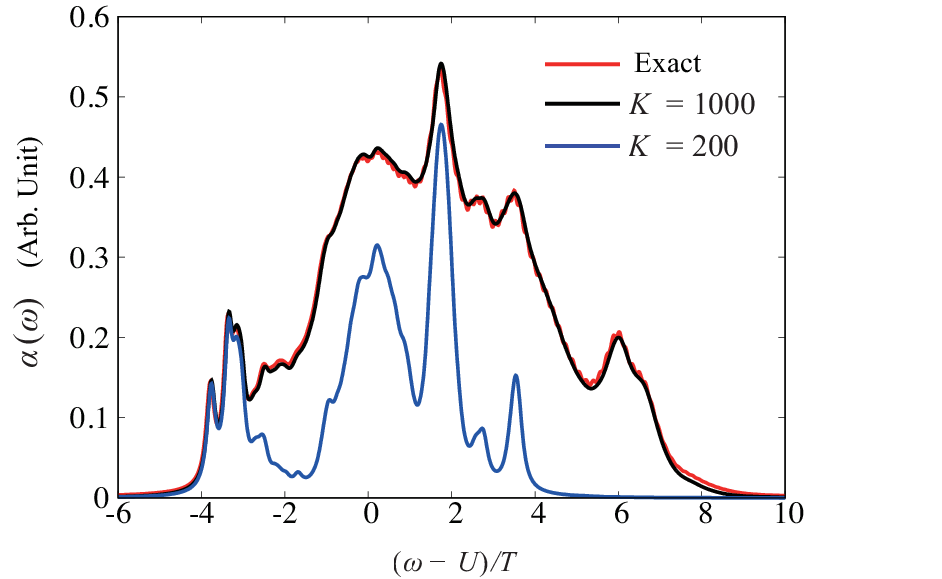}
  \caption{Comparison of the light absorption spectra calculated using the RSVD method with $K=200$ and $K=1000$ with the exact spectrum calculated using the Lanczos method for $T/U=0.1$. The broadening $\epsilon=0.12T$ is used.}
  \label{fig:Abs}
\end{figure}

We consider the $N=26$ cluster and the periodic boundary condition is used.
The shape of the cluster is shown as the tilted square that covers the bulk 2D lattice in Fig.~\ref{fig:lattice}.
We consider the case where the light field is polarized in the $x$-direction.
First, we present the light absorption spectrum calculated by the RSVD for $T/U=0.1$, which is shown in Fig.~\ref{fig:Abs}.
We use the parameters $N_{\rm t}=6000$ and $\Delta t T=0.1$. Then, $T_{\rm I}$ is large enough to ensure that there is only one dominant peak in ${\hat U}_{k}(\omega)$ and that the hybridization of the other energy eigenstates is negligible, except for modes with very small singular values.
The light absorption spectrum $\alpha (\omega)$ obtained via the RSVD with $K=1000$ is almost indistinguishable from the rigorous spectrum obtained using the Lanczos method.
 These results show that a practically exact $\alpha (\omega)$ can be reproduced using 1000 energy eigenstates in the $1.7 \times 10^7$-dimension Hilbert space. Use of the RSVD allows us to extract all energy eigenstates $|\phi _k \rangle$ that contribute to the light absorption spectrum (i.e., the one-photon active energy eigenstates). The results also show that $\Delta t T=0.1$ is small enough to allow the high-energy part of $\alpha (\omega)$ to be reproduced almost exactly.

\begin{figure}[thbp]\centering
 \includegraphics[width=120mm]{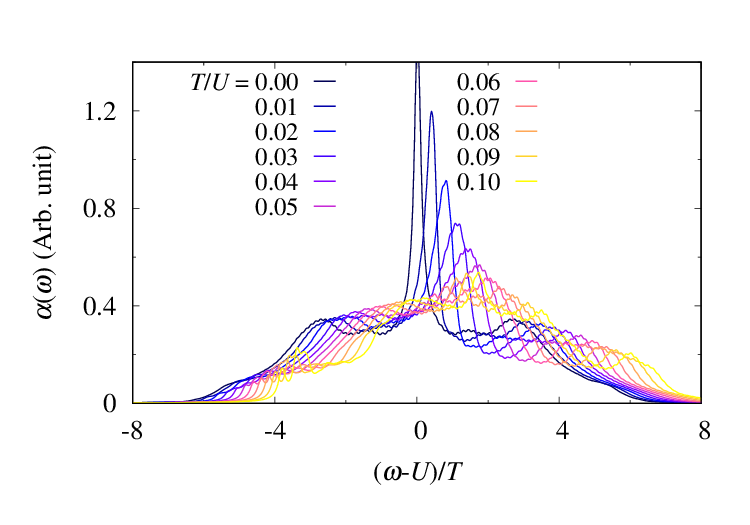}
  \caption{Comparison of the light absorption spectra for various values of $T/U$. The broadening $\epsilon=0.12T$ was used here.}
  \label{fig:AbsTU}
\end{figure}

The light absorption spectra for the 2D Mott insulators are essentially different to those for the 1D insulators. In the 1D Mott insulators, $\alpha (\omega)$ mainly consists of $N/2-1$ discrete peaks in the site $N$ cluster. Furthermore, only the energy gap increases with increasing $U$, and the spectrum shape remains almost unchanged when $T/U$ is varied within the strong correlation range $T/U \le 0.1$.\cite{CM} These characteristic properties come from the separation of the spin and charge degrees of freedom.\cite{CM}
In the 2D insulators, many more energy eigenstates make significant contributions to $\alpha (\omega)$, and an absorption band is formed with even a very small broadening of $\epsilon=0.12T$. Furthermore, the spectrum shape is strongly dependent on $T/U$ in 2D Mott insulators, as illustrated in Fig.~\ref{fig:AbsTU}. 

\begin{figure}[thbp]\centering
  \includegraphics[width=150mm,height=120mm]{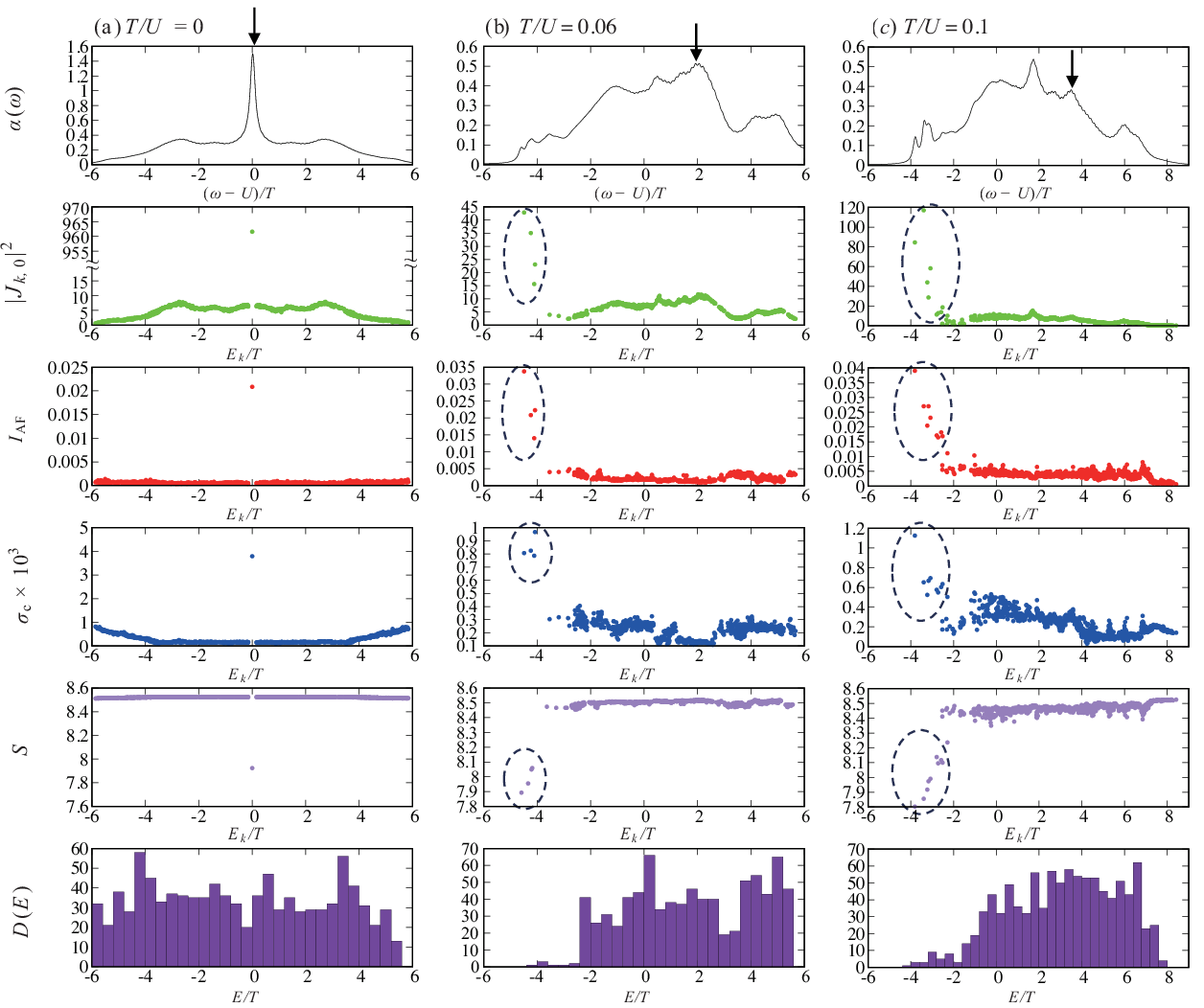}
  \caption{(a) Light absorption spectrum $\alpha (\omega)$ is shown as the solid line, and the central peak is indicated by the arrow for $T/U=0$. $|J_{k,0}|^2$, $I_{\rm AF}$, $\sigma_{\rm C}$, and $S$ for $|\phi _k \rangle$ are represented by the green, red, blue, and purple circles, respectively, as functions of $E_k$ for $T/U=0$. A histogram for the density of states $D(E)$ with a bin size of 0.4$T$ for $T/U=0$ is also shown. (b) Corresponding results for $T/U=0.06$. (c) Corresponding results for $T/U=0.1$. The points corresponding to the localized H-D states are circled using dotted lines in each case.}
  \label{fig:IAF}
\end{figure}

To understand the characteristic features of $\alpha (\omega)$ for 2D Mott insulators, we calculate all the one-photon active energy eigenstates by performing RSVD and investigate the physical properties of these eigenstates by analyzing the following physical quantities.
Because there is one H-D pair in an energy eigenstate $|\phi _k \rangle$, the correlation between the H and D charges $\xi(\bm{r})$ for $|\phi _k \rangle$ is given by:
\begin{eqnarray}
\xi(\bm{r}_m-\bm{r}_n) = 
\begin{cases}
\langle \phi _k | (1-n_n) (1-n_m) |\phi _k \rangle  & {\rm for}\quad n \neq m \\
0                                                   & {\rm for}\quad n = m
\end{cases}
\label{eq:CC}
\end{eqnarray}
where $1-n_n$ is the charge density operator at a site $n$, and $n_n=\sum_{\sigma} c_{n, \sigma}^{\dagger} c_{n, \sigma}$.
The D density at the site $\bm{r}$ when H exists at the site $(0,0)$ is given by $-(N/2)\xi(\bm{r})$, and the standard deviation of the spatial D distribution is then given by:
\begin{eqnarray}
\sigma_{\rm C}=\sqrt {\frac{1}{(N-1)}\sum_m^{\bm{r}_m \ne (0,0)}(-\frac{N}{2}\xi(\bm{r}_m)-\frac{1}{(N-1)})^2}.
\end{eqnarray}
The spin correlation $\eta(\bm{r})$ for $|\phi _k \rangle$ is defined by:
\begin{eqnarray}
\eta(\bm{r}_m-\bm{r}_n) &=& \langle \phi _k | \bm{S}_n \cdot \bm{S}_m |\phi _k \rangle,  \label{eq:SC}
\end{eqnarray}
where $\bm{S}_n$ is the spin operator at site $n$.
The spin structure factor $S(\pi,\pi)$, which is defined by
\begin{eqnarray}
S(\bm{q})=\sum_m \eta(\bm{r}_m) \exp[i\bm{q} \cdot \bm{r}_m],  \label{eq:SSF}
\end{eqnarray}
shows the magnitude of the AF spin correlation.
Because $\eta(0)$ is constant for the states that belong to $S_1$, we use
\begin{eqnarray}  \label{eq:IAF}
I_{\rm AF}=N^{-1}(S(\pi,\pi)-\eta(0)),
\end{eqnarray}
as the quantity that represents the magnitude of the AF spin order. 
We also consider the entanglement entropy here. We divide the entire system into two parts: subsystem $A$, which is shown in Fig.~\ref{fig:lattice}, and environment $B$. The partial trace ${\rm{Tr}}_B$ of an operator $\hat O$ is defined as ${\rm{Tr}}_B[\hat O]=\sum_n \langle n | \hat O |n \rangle$, where $|n \rangle$ represents the complete and orthonormal basis states in $B$. The reduced density matrix $\rho_A$ for $A$ is defined as $\rho_A={\rm{Tr}}_B [\rho]$, where $\rho = |\phi _k \rangle \langle \phi _k |$ is the density matrix. The entanglement entropy is then defined by
\begin{eqnarray}
S=-{\rm{Tr}}_A[\rho_A \log(\rho_A)].
\end{eqnarray}
A more localized wave function $|\phi _k \rangle$ corresponds to a smaller entanglement entropy $S$. We can thus determine the extent to which the wave function is localized from $S$.
We plot these quantities as a function of $E_k$ in Figs.~\ref{fig:IAF}(a), (b) and (c). As seen from these figures, as $T/U$ changes, the physical properties of the one-photon active energy eigenstates changes significantly, resulting in a significant change in the light absorption spectrum.

First, we consider the strong correlation limit where $T/U =0$. In this case, $H_{\rm eff}^{(1)}=P_1 \hat{K} P_1$ holds, and there are no interaction terms between the spins in the effective Hamiltonian $H_{\rm eff}^{(1)}$. The term $P_1 \hat{K} P_1$ describes the transfer of an H and a D to the neighboring sites and this term is therefore given by the transfer operator ${\hat T}_{d_i}$ for an H and a D as:
\begin{eqnarray}
P_1 \hat{K} P_1=-T\sum_{i=1}^4{\hat T}_{d_i}, \label{eq:P1KP1}
\end{eqnarray}
where ${\hat T}_{d_i}$ is defined by
\begin{eqnarray}
{\hat T}_{d_i} = \sum_{n,\sigma} \sum_l P_l c_{n(d_i), \sigma}^{\dag} c_{n, \sigma} P_l, \label{eq:hatT1H}
\end{eqnarray}
Here, $n(d_i)$ is the neighboring site of $n$ in the $d_i$-direction, and $d_1=x$, $d_2=y$, $d_3=-x$, and $d_4=-y$.
In this case, an H (D) is transferred to the neighboring site in the $-d_i$ ($d_i$) direction and one spin is transferred in the opposite direction by the operator ${\hat T}_{d_i}$.
As a result of the spin scattering induced by ${\hat T}_{d_i}$, solving for the eigenstates of $P_1 \hat{K} P_1$ is no longer a simple two-body problem involving an H and a D.

Before we investigate the energy eigenstates of $P_1 \hat{K} P_1$ in the half-filled case, we begin by considering simpler cases.
First, we consider the energy eigenstates in the Hilbert space $S_0$ when the electron number is $N-1$. These states have one H and no D, and the effective Hamiltonian of the Hubbard model for these states is then given by $P_0 \hat{K} P_0=-T\sum_{i=1}^4{\hat T}_{d_i}$.

Because ${\hat T}_{d_i}$ is translationally invariant, an eigenstate of ${\hat T}_{d_i}$ is given by:
\begin{eqnarray}
|\Psi _{{\bm q}^{\rm (H)},s} \rangle={\frac{1}{\sqrt {N} }}\sum_n \exp (i{\bm q}^{\rm (H)} \cdot {\bm r}_n) |f_s (n) \rangle, \label{eq:EE1H}
\end{eqnarray}
where the wave vector ${\bm q}^{\rm (H)}$ for an H is compatible with the $N=26$ cluster with the periodic boundary condition, and $|f_s (n) \rangle$ represents the normalized spin wave function of the spin state $s$.\cite{1H0D} The wave function has an H at site $n$, and this function is given by:
\begin{eqnarray}
|f_s (n) \rangle=\sum_{\sigma_2,\sigma_3,\cdots,\sigma_N} f_s (\sigma_2,\sigma_3,\cdots,\sigma_N) 
c_{n+1,\sigma_2}^{\dagger}c_{n+2,\sigma_3}^{\dagger}\cdots c_{N,\sigma_{N-n+1}}^{\dagger}c_{1,\sigma_{N-n+2}}^{\dagger}
\cdots c_{n-1,\sigma_N}^{\dagger}|{\rm vac}\rangle, \label{eq:SF1H}
\end{eqnarray}
where $|{\rm vac}\rangle$ is the vacuum state. Because we use a relative coordinate representation here, $f_s (\sigma_2,\sigma_3,\cdots,\sigma_N)$ is not dependent on $n$.
For $|\Psi _{{\bm q}^{\rm (H)},s} \rangle$ to be an eigenstate of ${\hat T}_{d_i}$, the wave function $|f_s (n) \rangle$ must satisfy the following equation:
\begin{eqnarray}
{\hat T}_{d_i}|f_s (n) \rangle=\exp[i\Delta q_{d_i}(s)]|f_s (n(-{d_i})) \rangle, \label{eq:TdSF1H}
\end{eqnarray}
where $\Delta q_{d_i}(s)$ is a real number that is dependent on $s$, and we use the fact that ${\hat T}_{d_i}$ is a unitary operator in the Hilbert space $S_0$ here.
From Eqs.~(\ref{eq:EE1H}) and (\ref{eq:TdSF1H}), we see that $|\Psi _{{\bm q}^{\rm (H)},s} \rangle$ is an eigenstate of ${\hat T}_{d_i}$ with an eigenvalue $\exp[i(q_{d_i}^{\rm (H)}+\Delta q_{d_i}(s))]$, where $q_{d_i}^{\rm (H)}$ is the $d_i$ component of ${\bm q}^{\rm (H)}$. Because the condition that ${\hat T}_{d_i}^{\dagger}={\hat T}_{-d_i}$ holds, $|\Psi _{{\bm q}^{\rm (H)},s} \rangle$ is also an energy eigenstate of $P_0 \hat{K} P_0$ with an energy eigenvalue of $E_{{\bm q}^{\rm (H)},s}=-2T\sum_{i=1}^2 \cos(q_{d_i}^{\rm (H)}+\Delta q_{d_i}(s))$.
The quantity $\Delta q_{d_i}(s)$ can be regarded as the phase shift induced by spin scattering. Additionally, $\Delta q_{d_i}(s)$ is a small quantity of the order of $1/N$ because only one spin is transferred by the operator ${\hat T}_{d_i}$. Because spin scattering does not occur and the case where $\Delta q_{d_i}(s)=0$ holds only occurs in the ferromagnetic state, the ground state is the ferromagnetic state with ${\bm q}^{\rm (H)}=0$. This state is called Nagaoka ferromagnetism.\cite{NagaokaF}

Next, we consider the energy eigenstates in the Hilbert space $S_1$ when the electron number is $N+1$. These states have no H and one D, and the effective Hamiltonian of the Hubbard model for these states is given by $P_1 \hat{K} P_1=-T\sum_{i=1}^4{\hat T}_{d_i}$.
Because the operator ${\hat T}_{d_i}$ is translationally invariant, an eigenstate of ${\hat T}_{d_i}$ is given by:
\begin{eqnarray}
|\Psi _{{\bm q}^{\rm (D)},s} \rangle={\frac{1}{\sqrt {N} }}\sum_n \exp (i{\bm q}^{\rm (D)} \cdot {\bm r}_n) |f_s (n) \rangle, \label{eq:EE1D}
\end{eqnarray}
where the wave vector ${\bm q}^{\rm (D)}$ of a D is compatible with the $N=26$ cluster with the periodic boundary condition, and $|f_s (n) \rangle$ represents the normalized spin wave function of the spin state $s$. The wave function has a D at site $n$, and this function is given by:
\begin{eqnarray}
|f_s (n) \rangle=\sum_{\sigma_2,\sigma_3,\cdots,\sigma_N} f_s (\sigma_2,\sigma_3,\cdots,\sigma_N)
c_{n,\uparrow}^{\dagger} c_{n,\downarrow}^{\dagger}
c_{n+1,\sigma_2}^{\dagger}c_{n+2,\sigma_3}^{\dagger}\cdots c_{N,\sigma_{N-n+1}}^{\dagger}c_{1,\sigma_{N-n+2}}^{\dagger}
\cdots c_{n-1,\sigma_N}^{\dagger}|{\rm vac}\rangle, \label{eq:SF1D}
\end{eqnarray}
The normalized spin wave function satisfies the condition that:
\begin{eqnarray}
{\hat T}_{d_i}|f_s (n) \rangle=\exp[-i\Delta q_{d_i}(s)]|f_s (n(d_i)) \rangle, \label{eq:TdSF1D}
\end{eqnarray}
and $|\Psi _{{\bm q}^{\rm (D)},s} \rangle$ is an eigenstate of ${\hat T}_{d_i}$ with an eigenvalue $\exp[-i(q_{d_i}^{\rm (D)}+\Delta q_{d_i}(s))]$.
Note here that the shift $\Delta q_{d_i}(s)$ is the same in the cases with only one H and only one D because the effects of the spin scattering caused by the charge transfer on the spin states are the same in these two cases.
Because the condition ${\hat T}_{d_i}^{\dagger}={\hat T}_{-d_i}$ holds, $|\Psi _{{\bm q}^{\rm (D)},s} \rangle$ is also an energy eigenstate of $P_1 \hat{K} P_1$ with an energy eigenvalue $E_{{\bm q}^{\rm (D)},s}=-2T\sum_{i=1}^2 \cos(q_{d_i}^{\rm (D)}+\Delta q_{d_i}(s))$.

Next, we consider the energy eigenstates of the effective Hamiltonian $P_1 \hat{K} P_1=-T\sum_{i=1}^4{\hat T}_{d_i}$ in the half-filled case, which has one H-D pair.
In the first step, we temporarily neglect the constraint that an H and a D cannot occupy the same site.
We then consider the following two approximations of the effect of the spin scattering:
(i) the interaction between an H and a D induced by the spin scattering is negligible, and (ii) the transfer of an H or a D only induces the phase shift in the spin wave function, and this shift is the same as that in the case of one H and no D and the case of no H and one D. Using these approximations, the eigenstates of ${\hat T}_{d_i}$ are then given by:
\begin{eqnarray}
|\Psi _{{\bm q}^{\rm (H)},{\bm q}^{\rm (D)},s} \rangle={\frac{1}{N}}\sum_{n,m} \exp (i{\bm q}^{\rm (H)} \cdot {\bm r}_{n}+i{\bm q}^{\rm (D)} \cdot {\bm r}_m) |f_s (n,m) \rangle,\label{eq:EEF1D1HnoC}
\end{eqnarray}
where $|f_s (n,m) \rangle$ is the spin wave function with an H at site $n$ and a D at site $m$ that satisfies the following relation:
\begin{eqnarray}
{\hat T}_{d_i}|f_s (n,m) \rangle=\exp[i\Delta q_{d_i}(s)]|f_s (n(-d_i),m) \rangle + \exp[-i\Delta q_{d_i}(s)]|f_s (n,m(d_i)) \rangle, \label{eq:TdSF1H1D}
\end{eqnarray}
and the eigenvalue of $|\Psi _{{\bm q}^{\rm (H)},{\bm q}^{\rm (D)},s} \rangle$ is $\exp[i(q_{d_i}^{\rm (H)}+\Delta q_{d_i}(s))]+\exp[-i(q_{d_i}^{\rm (D)}+\Delta q_{d_i}(s))]$.
$|\Psi _{{\bm q}^{\rm (H)},{\bm q}^{\rm (D)},s} \rangle$ is also an energy eigenstate of $P_1 \hat{K} P_1$ and has the energy eigenvalue $-2T\sum_{i=1}^2 \{\cos(q_{d_i}^{\rm (H)}+\Delta q_{d_i}(s))+\cos(q_{d_i}^{\rm (D)}+\Delta q_{d_i}(s))\}$.

The constraint is satisfied by considering a linear combination of two degenerate eigenstates:  ${\frac{1}{\sqrt{2}i}}\{|\Psi _{{\bm q},-{\bm q},s} \rangle-|\Psi _{-{\bm q},{\bm q},s} \rangle \}$. Consequently, the energy eigenstates of $P_1 \hat{K} P_1$ are given by:
\begin{eqnarray}
|\psi _{{\bm q},s} \rangle={\frac{\sqrt{2}}{N}}\sum_{n,m} \sin [{\bm q} \cdot ({\bm r}_{n}-{\bm r}_m)] |f_s (n,m) \rangle, \label{eq:EEF1D1H}
\end{eqnarray}
and the energy eigenvalue of $|\psi _{{\bm q},s} \rangle$ is given by
\begin{eqnarray}
E_{{\bm q},s}=-2T\sum_{i=1}^2 \{\cos(q_{d_i}+\Delta q_{d_i}(s))+\cos(q_{d_i}-\Delta q_{d_i}(s))\}. \label{eq:EEV1D1H}
\end{eqnarray}
The H-D wave vector ${\bm q}$ must be compatible with both the periodic boundary condition and the constraint that an H and a D cannot occupy the same site. The latter constraint cannot be satisfied for the wave vectors $(0,0)$, $(0,\pi)$, $(\pi,0)$, and $(\pi,\pi)$ because $|\Psi _{{\bm q},-{\bm q},s} \rangle$ and $|\Psi _{-{\bm q},{\bm q},s} \rangle$ are the same state in each of these cases.

The H-D model describes the transfer of an H and a D under the constraint that an H and a D cannot occupy the same site. The spin degrees of freedom are not included in the H-D model.
The charge distribution of $|\psi _{{\bm q},s} \rangle$ is the same as that for the energy eigenstate of the H-D model when using the same H-D wave vector ${\bm q}$. In this paper, $|\psi _{{\bm q},s} \rangle$ are therefore regarded as free H-D states. Note, however, that the spin scattering induces a phase shift and the spin-charge coupling is not negligible, even within the strong interaction limit $T/U=0$.

Next, we consider the spin structure of these free H-D states. A free H-D state $|\psi _{{\bm q},s} \rangle$ is given by a linear combination of the spin wave functions $|f_s (n,m) \rangle$, and all these spin wave functions are coupled through the transfer operator ${\hat T}_{d_i}$, as shown by Eq.~(\ref{eq:TdSF1H1D}), except for the case where $|q_x|=|q_y|=\pi/2$. When $|q_x|=|q_y|=\pi/2$ holds, if $|f_s (n,m) \rangle$ has a finite weight, then the weights of $|f_s (n(d_i),m) \rangle$ and $|f_s (n,m(d_i)) \rangle$ are both zero in $|\psi _{{\bm q},s} \rangle$, as shown by Eq.~(\ref{eq:EEF1D1H}). Therefore, the spin wave functions $|f_s (n,m) \rangle$ that construct $|\psi _{{\bm q},s} \rangle$ are not coupled through ${\hat T}_{d_i}$ in this case alone.

\begin{figure}[thbp]\centering
 \includegraphics[width=80mm]{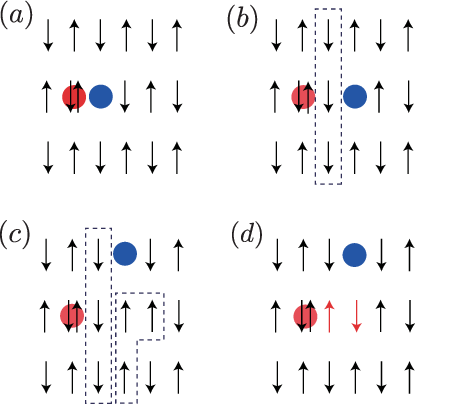}
  \caption{(a) Spin configuration of the AF spin state $|f_s (n,m) \rangle$ with an H-D pair located at nearest-neighbor sites. (b) Spin configuration of $|f_s (n(x),m) \rangle$. (c) Spin configuration of $|f_s (n(x,y),m) \rangle$, where $n(x,y)$ is the site neighboring $n(x)$ in the $y$-direction. (d) Spin configuration of the state where a spin flip of an antiparallel pair indicated by the red arrows is induced in $|f_s (n(x,y),m) \rangle$. In all parts of the figure, the up (down) spins are shown by the up (down) arrows, the H and the D are represented by the blue and red circles, respectively, and the parallel spin pairs induced by the H transfer are surrounded by dotted lines.}
  \label{fig:AFHT}
\end{figure}

If $|f_s (n,m) \rangle$ has the AF spin order, this order is destroyed locally around the site $n$ ($m$) in $|f_s (n(d_i),m) \rangle$ ($|f_s (n,m(d_i)) \rangle$) by the spin scattering induced by the transfer of an H (D), and the number of parallel spin pairs increases when an H (D) is transferred further, as depicted schematically in Fig.~\ref{fig:AFHT}(a), (b) and (c). If all the spin wave functions in a free H-D state $|\psi _{{\bm q},s} \rangle$ are coupled through ${\hat T}_{d_i}$, then the AF spin order is destroyed by the free translation of an H (D). Therefore, $|\psi _{{\bm q},s} \rangle$ can have the AF spin order only when $|q_x|=|q_y|=\pi/2$.

Because spin scattering does not occur in the ferromagnetic state, free H-D states with a ferromagnetic order must exist. However, because the ground state has the AF spin order, the transition dipole moments to the free H-D states with the ferromagnetic spin order are effectively zero. Therefore, the one-photon active free H-D states have neither AF nor ferromagnetic spin orders, except for the case where $|q_x|=|q_y|=\pi/2$.

\begin{figure}[thbp]\centering
  \includegraphics[width=120mm]{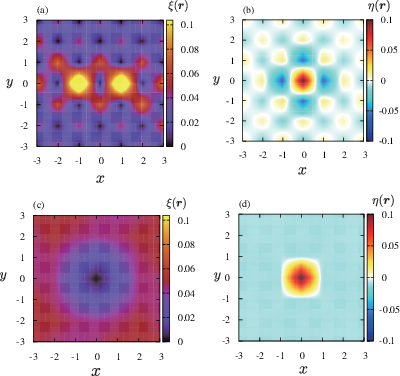}
  \caption{(a) $\xi(\bm{r})$ and (b) $\eta(\bm{r})$ for $|\phi _k \rangle$ at $E_k = 0$, and (c) $\xi(\bm{r})$ and (d) $\eta(\bm{r})$ for $|\phi _k \rangle$ at $E_k = 5.85T$ within the strong correlation limit $T/U=0$.}
  \label{fig:SCCF0}
\end{figure}

The spin and charge correlation function characteristics of the free H-D states $|\psi _{{\bm q},s} \rangle$ agree well with those of the one-photon active energy eigenstates that were obtained numerically via RSVD. As shown in Fig.~\ref{fig:IAF}(a), the magnitude of the AF spin order $I_{\rm AF}$ for the energy eigenstate with $E_k = 0$ is much higher than the magnitudes for the other states. We show the spin correlation function $\eta(\bm{r})$ for the $|\phi _k \rangle$ at $E_k = 0$ and $5.85T$ in Fig.~\ref{fig:SCCF0}(b) and (d), respectively. We see that $|\phi _k \rangle$ with $E_k = 0$ has the AF spin order. Furthermore, the values of $|\eta(\bm{r})|$ for $\bm{r} \ne 0$ are very small, and no spin order exists in $|\phi _k \rangle$ when $E_k = 5.85T$. We see from the results of the numerical calculations that the spin correlations are very weak in the other one-photon active energy eigenstates, except for the case where $E_k = 0$.
As shown in Eq.~(\ref{eq:EEV1D1H}), $E_{{\bm q},s}=0$ holds irrespective of the spin state $s$ when $|q_x|=|q_y|=\pi/2$, and many of the free H-D states are degenerate at $E_k =0$. This property of the free H-D states is also consistent with the numerical result that only the single energy eigenstate has the strong AF spin order.

The charge correlation for a free H-D state $|\psi _{{\bm q},s} \rangle$ is given by
\begin{eqnarray}
\xi(\bm{r})=-{\frac{4}{N^2}}\sin ^2[{\bm q} \cdot {\bm r}]. \label{eq:CCFEEF1D1H}
\end{eqnarray}
In the case where $|q_x|=|q_y|=\pi/2$, $\xi(\bm{r}_m-\bm{r}_n)=0$ holds when both $n$ and $m$ belong to the same bipartite sublattice; this shows that if an H is on one bipartite sublattice, then a D is always on the other bipartite sublattice. As the values of $q_x$ and $q_y$ become closer to 0 ($\pi$) and $E _{{\bm q},s}$ approaches the low-energy (high-energy) band edge, the spatial variations in the charge correlations become more gradual.

We show $\xi(\bm{r})$ for $|\phi _k \rangle$ with values of $E_k =0$ and $5.85T$ in Fig.~\ref{fig:SCCF0}(a) and (c), respectively. The characteristic charge correlation function of the free H-D state for the case where $|q_x|=|q_y|=\pi/2$ is shown, i.e., $\xi(\bm{r}_m-\bm{r}_n)=0$ holds when $n$ and $m$ belong to the same bipartite sublattice in $|\phi _k \rangle$ when $E_k =0$. Because $|\phi _k \rangle$ with $E_k =0$ is given by a linear combination of the degenerate free H-D states with $|q_x|=|q_y|=\pi/2$ and different signs of $q_x$ and $q_y$, and spin states $s$, the value of $\xi(\bm{r}_m-\bm{r}_n)$ is not constant when $n$ and $m$ belong to different bipartite sublattices.
In the case of $|\phi _k \rangle$ with $E_k = 5.85T$, which is near the high-energy band edge, the Fourier transformation of the charge correlation function $\xi(\bm{r})$ shows a peak around the wavelength that is comparable to $\sqrt{N}$. We have also investigated $\xi(\bm{r})$ for the other one-photon active energy eigenstates. Because of the degeneracy of $|\psi _{{\bm q},s} \rangle$, the value of $\xi(\bm{r})$ given by Eq.~(\ref{eq:CCFEEF1D1H}) is not reproduced in the numerically obtained $|\phi _k \rangle$, but we confirmed that the spatial variation in $\xi(\bm{r})$ becomes more gradual as $E_k$ becomes closer to the band edges. 
The characteristic spin and charge structures of the numerically obtained one-photon active energy eigenstates agree very well with the corresponding quantities for the free H-D states.

The light absorption spectrum $\alpha (\omega)$ for $T/U = 0$ consists of a broad band with a sharp peak at $\omega-U=0$. This broad band is symmetrical about $\omega-U=0$ and extends from $U-6T$ to $U+6T$, and $\alpha (\omega)$ decreases as $\omega$ approaches the band edges. These characteristic features of $\alpha (\omega)$ can be explained using the physical properties of the free H-D states.

In the H-D model, the energy eigenvalue is determined from the H-D wave vector ${\bm q}$. Because the number of ${\bm q}$ that is compatible with both the periodic boundary condition and the constraint that an H and a D cannot occupy the same site is smaller than $N$, $\alpha (\omega)$ consists of only a few isolated peaks in the small-sized cluster.
In the model presented here, the energy eigenvalues of the free H-D state are determined not only from ${\bm q}$ but also from the phase shift $\Delta {\bm q}(s)$, which arises from the spin scattering and is dependent on the spin state $s$. As a result of the large number of spin degrees of freedom, approximately 1000 energy eigenstates make nonnegligible contributions to $\alpha (\omega)$, and the broad band is thus formed.
In the 1D Hubbard model, as a result of the spin-charge separation, the energy eigenvalues of the one-photon excited states are mainly determined from the wave number $q$ of an H and a D, and $\alpha (\omega)$ then mainly consists of a few isolated peaks.\cite{CM}
The broad band structure in the 2D effective model originates from spin-charge coupling.

A logarithmic singularity has been reported  in the density of states for noninteracting electrons on a 2D square lattice.\cite{DOS1,DOS2} However, we can see that $\alpha (\omega)$ is mainly determined from $|J_{k,0}|^2$ by comparing $\alpha (\omega)$ with $|J_{k,0}|^2$, as shown in Fig.~\ref{fig:IAF} (a). As Eq.~(\ref{eq:TDM}) shows, $J_{k,0}$ is given by the overlap between $|\phi _k \rangle$ and ${\hat J}|\phi _0 \rangle$. In ${\hat J}|\phi _0 \rangle$, an H-D pair is generated at neighboring sites along the $x$-direction in the AF spin background, as indicated by Eq.~(\ref{eq:J}).
Therefore, $|J_{k,0}|$ tends to be higher for $|\phi _k \rangle$ with a higher $I_{\rm AF}$ and a higher $|\xi((1,0))|$, which is proportional to the probability that an H and a D exist at neighboring sites along the $x$-direction.

The transition dipole moment is much higher for $|\phi _k \rangle$ with $E_k = 0$ than for any of the other states, and this is mainly the result of $I_{\rm AF}$ being much larger for this state, which is characteristic of free H-D states with $|q_x|=|q_y|=\pi/2$.
The sharp peak observed at $\omega-U=0$ occurs because of the excitation to the single energy eigenstate with the dominant transition dipole moment, which stems from the properties of the free H-D states.

Next, we consider the origin of the broad band structure. Because the values of $I_{\rm AF}$ for the one-photon active energy eigenstates are small and are nearly constant, $|J_{k,0}|$ is determined mainly from $|\xi((1,0))|$, except for the case where $E_k=0$. In the free H-D state $|\phi _{{\bm q},s} \rangle$, the relation that $|\xi((1,0))| = {\frac{4}{N^2}} \sin^2(q_x)$ holds.
For the two wave vectors ${\bm q}^{\pm}=(\pi/2,\pi/2) \pm {\bm q}^{\prime}$, $E_{{\bm q}^+,s}=-E_{{\bm q}^-,s}$ holds , and the value of $|\xi((1,0))|$ for $|\psi _{{\bm q}^+,s} \rangle$ and that for $|\psi _{{\bm q}^-,s} \rangle$ are equal.
As ${\bm q}$ approaches $(0,0)$ ($(\pi,\pi)$), $E _{{\bm q},s}$ approaches the upper (lower) band edge, and $\xi((1,0))$ approaches zero.
The value of $|\xi((1,0))|$ is highest at $q_x=\pm \pi/2$, and $|\xi((1,0))|$ can reach its highest value only when $-4T \le E _{{\bm q},s} \le 4T$.
These characteristics of the free H-D states are consistent with the numerically obtained results that indicated that the $E_k$ dependence of $|J_{k,0}|$ is symmetrical about $E_k=0$, that $|J_{k,0}|$ is nearly constant for $-3T \lesssim E _k-U \lesssim 3T$, and that $|J_{k,0}|$ decreases with decreasing (increasing) $E_k$ for $E_k \lesssim -3T$ ($E_k \gtrsim 3T$) and approaches zero near the low-energy (high-energy) band edge.
The characteristics of $\alpha (\omega)$ for $T/U = 0$ can be explained using the charge and spin structures of the free H-D energy eigenstates.
The density of the free H-D states $D(E)$ decreases as $E$ approaches the band edges. This also contributes to the reduction in $\alpha (\omega)$ near the band edges.

As shown in Fig.~\ref{fig:AbsTU}, as $T/U$ increases from $T/U=0$, the sharp peak at the center of the band broadens and blue shifts, and $\alpha(\omega)$ for $\omega < \omega_{\rm CP}$ ($\omega > \omega_{\rm CP}$) increases (decreases), where $\omega_{\rm CP}$ is the central peak energy.
Furthermore, new peaks that are separated from the broad band toward the lower energy, as seen from $D(E)$, appear when $T/U \gtrsim 0.03$, and these peaks become higher as $T/U$ increases. As a result, the shape of $\alpha (\omega)$ for a realistic value for two-dimensional Mott insulators of $T/U=0.1$ is essentially different from that at the strong correlation limit of $T/U=0$. This result is in contrast to the 1D case, where the spectrum shape remains almost unchanged when $T/U$ is changed under the condition that $T/U \le 0.1$.

To understand the characteristics of $\alpha (\omega)$ for $T/U=0.1$, we investigated the physical properties of the one-photon active energy eigenstates.
As shown in Fig.~\ref{fig:IAF}(c), the one-photon active energy eigenstates can be classified into two categories.
The magnitudes of the transition dipole moments $|J_{k,0}|$, the magnitudes of the AF spin order $I_{\rm AF}$, and the standard deviations of the spatial H and D distribution $\sigma_{\rm C}$ of the nine energy eigenstates are significantly higher, and the entanglement entropies $S$ of these eigenstates are significantly smaller than those of the other states.

\begin{figure}[thbp]\centering
  \includegraphics[width=120mm]{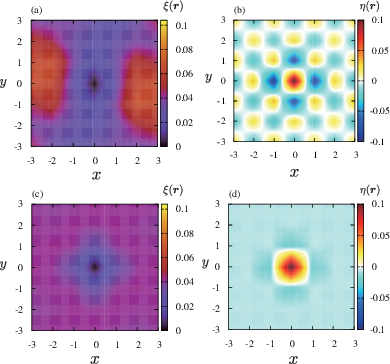}
  \caption{(a) $\xi(\bm{r})$ and (b) $\eta(\bm{r})$ for $|\phi _k \rangle$ at $E_k= -3.82T$, and (c) $\xi(\bm{r})$ and (d) $\eta(\bm{r})$ for $|\phi _k \rangle$ at $E_k=1.69T$ for $T/U=0.1$.}
  \label{fig:SCCF01}
\end{figure}

We compare $\xi(\bm{r})$ and $\eta(\bm{r})$ for $|\phi _k \rangle$ with $E_k=-3.82T$ (one of the nine energy eigenstates) with the corresponding quantities for $E_k=1.69T$ (one of the other eigenstates) in Fig.~\ref{fig:SCCF01}. $|\phi _k \rangle$ with $E_k=1.69T$ shows the typical charge and spin structure of a free H-D state. The AF spin order is much stronger, and $|\xi(\bm{r})|$, which is proportional to the D density when an H exists at site $(0,0)$, is much more localized around sites $(2,-1)$ and $(-2,1)$ in the case of $|\phi _k \rangle$ with $E_k=-3.82T$ than in the case of $|\phi _k \rangle$ with $E_k=1.69T$. The much higher values of $\sigma_{\rm C}$ for the nine energy eigenstates result from the more localized H and D distributions.
We cannot determine whether the H and the D are localized or not from the long-range behavior of $\xi(\bm{r})$ in the finite cluster calculations in the present case. However, because the AF spin order is destroyed by the transfer of an H and a D, the strong AF spin order indicates localization of an H and a D.
Furthermore, the entanglement entropy $S$ and the density of states $D(E)$ are significantly smaller within the energy regions of the nine energy eigenstates than in the other energy regions. These results are consistent with the conclusion that an H and a D are localized within these energy eigenstates.
These nine energy eigenstates are therefore described as localized H-D states.

An H and a D are localized around $r_{\rm H-D}=\sqrt {5}$, where $r_{\rm H-D}$ is the distance between the H and the D, in $|\phi _k \rangle$ with $E_k=-3.82T$. We have also investigated $\xi(\bm{r})$ for the other localized states. An H and a D are localized around the distance $r_{\rm H-D}=\sqrt {5}$ in cases where $E_k=-3.39T$, $-3.22T$, and $-3.08T$, and an H and a D are also localized around the distance $r_{\rm H-D}=\sqrt {13}$ in the cases where $E_k=-2.79T$, $-2.72T$, $-2.58T$, and $-2.51T$. In addition, in the case where $E_{k}=-3.17T$, an H and a D are localized around both $r_{\rm H-D}=\sqrt{5}$ and $\sqrt{13}$. These characteristics are in contrast to the H-D exciton state that exists when the Coulomb interactions between the different sites are strong enough. In the lowest energy H-D exciton state, $|\xi(\bm{r})|$ is a decreasing function of $r_{\rm H-D}$.\cite{effM2}

This characteristic charge distribution can be understood as follows.
The one-photon active energy eigenstates have a nonzero overlap with the solution to the time-dependent Schr\"{o}dinger equation with the initial state $|\psi (0) \rangle = {\hat J} |\phi _0 \rangle$, because the relationship $\langle \phi _k |\psi (t) \rangle = \exp(-iE_kt) J_{k,0}$ holds. There is an H-D pair at a nearest-neighbor site on the AF spin background in the initial state ${\hat J} |\phi _0 \rangle$. An H or a D is then transferred to the neighboring site by the transfer term $P_1 \hat{K} P_1$ in $H_{\rm eff}^{(1)}$, and the neighboring parallel spin pairs are generated by the transfer, as shown schematically in Fig.~\ref{fig:AFHT}(a), (b) and (c). A spin flip of an antiparallel pair is induced by the spin-spin interaction term in $H_{\rm eff}^{(1)}$. The parallel spin pairs that are generated by successive transfers of charge carriers can only be removed by the spin flip when an H and a D exist on different bipartite sublattices, i.e., when $r_{\rm H-D}=1$, $\sqrt {5}$, and $\sqrt {13}$ in the current $N=26$ cluster, as shown schematically in Fig.~\ref{fig:AFHT}(d).
Therefore, the energy gain due to the AF spin order that results from the localization of an H and a D reaches local maxima at these distances.

The AF spin order survives and the spin-spin interaction energy is much lower in the localized H-D states with these distances than in the free H-D states, where the AF spin order is destroyed. However, the H and D transfer energy is also increased by the localization. The increases in the spin-spin interaction energies from the AF ground state are roughly estimated to be $6J|\eta_0|$, $8J|\eta_0|$, and $8J|\eta_0|$ for the cases of $r_{\rm H-D}=1$, $\sqrt {5}$, and $\sqrt {13}$, respectively, where $\eta_0$ is the spin correlation between neighboring states in the AF ground state, and the increase is shown to be lowest for $r_{\rm H-D}=1$.
When a localized state with a distance $r_{\rm H-D}$ is represented by a linear combination of the free H-D states $|\psi _{{\bm q},s} \rangle$,
the states with the wave vectors of $|{\bm q}|\simeq\pi/r_{\rm H-D}$ have the dominant weights. Therefore, the transfer energy is much higher in the localized states with $r_{\rm H-D}=1$ than in the states with $r_{\rm H-D}=\sqrt {5}$ and $r_{\rm H-D}=\sqrt {13}$, and the difference is related to the order of $T$. This is the reason why an H and a D are localized with a distance of $r_{\rm H-D}=\sqrt {5}$ or $r_{\rm H-D}=\sqrt {13}$ in the localized H-D states.
Note here that the localization does not come from the direct Coulomb interaction; in fact, it comes from the spin-spin interaction.

Consequently, the one-photon active energy eigenstates can be classified into localized H-D states and free H-D states. The localized H-D energy eigenstates have much higher $J_{k,0}$ values than the free H-D energy eigenstates, and the density of states is much lower in the energy region of a localized H-D state than in the corresponding region of a free H-D state. As a result, $\alpha(\omega)$ consists of a few sharp peaks in the low-energy region of the localized H-D states for $-4T \lesssim \omega -U \lesssim -2.5T$. In contrast, the free H-D states form a broad band that extends from $-2.5T \lesssim \omega -U \lesssim 8T$. The peaks that result from the localized H-D states are separated toward the lower energy from the broad band that results from the free H-D states as seen from the dip in $D(E)$ shown in Fig.~\ref{fig:IAF}(c).

Next, we consider how $\alpha(\omega)$ and the photoexcited states change with changes in $T/U$.
The localized H-D states appear when $T/U \gtrsim 0.03$. There are four localized H-D states for $T/U = 0.03$. As $T/U$ is increased from $T/U=0.03$, the number of localized H-D states increases, the H and the D becomes more localized, and $I_{\rm AF}$ and $|J_{k,0}|$ both increase in the localized H-D states, as illustrated by comparison of Fig.~\ref{fig:IAF}(b) and (c). The interaction between the spins strengthens and the gain of the spin-spin interaction energy that results from the localization of the H and the D also increases as $T/U$ increases. This characteristic $T/U$ dependence of the one-photon active energy eigenstates can be attributed to the increased gain of the spin-spin interaction energy. The peaks caused by the localized H-D states become more prominent as $T/U$ increases because of the increase in $|J_{k,0}|$ and the increase in the number of these states. 
 
The sharp peak centered at $\omega -U=0$ becomes broader as $T/U$ is increased from zero. Numerical calculations show that the sharp peak results from excitation of a few nondegenerate energy eigenstates, and these energy eigenstates have the characteristic charge and spin structures of the free H-D states $|\psi _{{\bm q},s} \rangle$ with $|q_x|=|q_y|=\pi/2$ for $T/U = 0.01$. Therefore, the broadening that occurs in the extremely strong correlation region $T/U \lesssim 0.01$ is mainly because the degeneracy of $|\psi _{{\bm q},s} \rangle$ with $|q_x|=|q_y|=\pi/2$ is lifted by the introduction of spin-spin interactions. For the region where $T/U \gtrsim 0.03$, we cannot confirm the characteristic spin and charge structures of $|\psi _{{\bm q},s} \rangle$ with $|q_x|=|q_y|=\pi/2$ in the energy eigenstates that contribute to the peak. This shows that $|\psi _{{\bm q},s} \rangle$ with different wave vectors are strongly hybridized in these energy eigenstates, and this behavior leads to the broadening in the correlation region.

Next, we consider the broad band of the free H-D states.
When $T/U=0$, $\alpha(\omega)$ is nearly constant within the central part of the band for $-3T \lesssim \omega -U \lesssim 3T$ when the sharp central peak is neglected. As $T/U$ increases, both $I_{\rm AF}$ and $\sigma_{\rm C}$ also increase slightly in the free H-D states, showing that the translation motion of the H and the D is also slightly suppressed to gain the spin-spin interaction energy in these states. Furthermore, a smaller energy eigenvalue tends to correspond to a greater increase in $\sigma_{\rm C}$ and $I_{\rm AF}$, as shown in Fig.~\ref{fig:IAF}(b) and (c). The effect of the potential energy that results from the spin degrees of freedom increases as the kinetic energy of the H and the D decreases. As a result of the characteristic $E_k$ dependence of $I_{\rm AF}$, $\alpha(\omega)$ becomes approximately a decreasing function of $\omega$ in the central part of the band for $0 \lesssim \omega -U \lesssim 6T$ for $T/U=0.1$ when the small peaks are neglected.

The light absorption spectrum $\alpha(\omega)$ for the realistic magnitude of the Coulomb interaction $T/U=0.1$ agrees well with the experimental results for 2D Mott insulators composed of La$_2$CuO$_4$, Nd$_2$CuO$_4$, and Sr$_2$CuO$_2$Cl$_2$.\cite{2DabsE1,2DabsE2,SCCinDoped2DMI13} The characteristic $\alpha(\omega)$ of the 2D Mott insulator can be explained by considering the effects of the spin degrees of freedom on the transfer of an H and a D.

\section{Summary and discussions} \label{sec:SD}
We have proposed a new approach to extract the important degrees of freedom in quantum dynamics induced by an external stimulus by performing RSVD of the coefficient matrix $C$, and then applied this method to analyze the light absorption spectra in 2D Mott insulators.
We found that a nearly exact light absorption spectrum $\alpha(\omega)$ can be reproduced by using as few as 1000 energy eigenstates in $1.7 \times 10^7$-dimension Hilbert space. Furthermore, we calculated all these one-photon active energy eigenstates and analyzed their physical properties, which then enabled us to understand the characteristic $\alpha(\omega)$ of 2D Mott insulators.
For a realistic magnitude for the Coulomb interaction $T/U=0.1$, the one-photon active energy eigenstates were classified into free H-D and localized H-D states. In the free H-D states, the main effect of the spin degrees of freedom on the transfer of an H and a D was a phase shift, and the H and the D moved freely. In the localized H-D states, a H and a D were localized with the relative distances of $\sqrt {5}$ or $\sqrt {13}$, and the AF spin order was much stronger than the free H-D states. The localization of the charge carriers was caused by the interaction between them, which originated from the spin degrees of freedom.
Sharp peaks are caused by the excitation to the localized H-D states, and these peaks are separated from the broad band that results from the free H-D states toward the lower energy. The numerical results provide reasonable reproductions of the experimentally obtained optical conductivity characteristics in the 2D Mott insulators composed of La$_2$CuO$_4$, Nd$_2$CuO$_4$, and Sr$_2$CuO$_2$Cl$_2$.\cite{2DabsE1,2DabsE2,SCCinDoped2DMI13} In the experimental results, a peak was observed at the low-energy edge of the continuum band. We believe that this peak occurs because the peaks caused by the localized H-D state are not separated from the continuum band of the free H-D states because of the large broadening effect.

The H-D binding effect caused by the Coulomb interaction between the different sites is not investigated in this paper.
It has been shown that H-D exciton states exist when $V/T \gtrsim 2$, where $V$ is the Coulomb interaction energy between the neighboring sites, and the peaks caused by the excitations to the H-D exciton states are dominant within the light absorption spectrum in the effective 2D extended Hubbard model in the strong interaction case.\cite{effM2}
The absorption spectra for $V/T \gtrsim 2$ differ greatly from the experimentally obtained spectra in the 2D Mott insulators La$_2$CuO$_4$ and Nd$_2$CuO$_4$, but the experimental results can be reproduced reasonably well in the Hubbard model or in its effective model in the strong interaction case. We therefore consider the localized H-D states that result from the spin degrees of freedom to be dominant in the low-energy region of $\alpha(\omega)$ in these 2D Mott insulators.
The localized H-D states can be detected from the photon energy dependence of the photoconductivity.
The localized H-D states are not expected to contribute to the photoconductivity. Therefore, the photoconductivity rises above the band edge, i.e., at the boundary energy between the localized and free H-D states.

In this paper, we have shown that all the one-photon active energy eigenstates can be calculated using the proposed method based on RSVD that considers the linear response to the photoexcitation, and this enables us to understand the light absorption spectrum in the 2D Mott insulators. However, this method is not limited to analysis of the linear response. The proposed method can also be used to extract the important degrees of freedom for the dynamics when the photoexcited states change greatly, e.g., as in the photoinduced phase transitions.
Transient absorption spectroscopy represents a powerful experimental tool for investigation of the dynamics of photoinduced phase transitions. However, it is often difficult to understand the dynamics from the transient absorption spectrum, similar to the difficulty of understanding the linear absorption spectra of 2D Mott insulators. The proposed method can be used to analyze the transient absorption spectrum by performing RSVD of $C_{\rm probe}-C$, where $C_{\rm probe}$ is the coefficient matrix when the pump and probe pulses are irradiated and $C$ is the corresponding matrix when only the pump pulse is irradiated. Although the modes obtained are not the energy eigenstates because of the uncertainty relation between the time and the energy in this case, analysis of these important modes will still advance our understanding of the optical transient absorption spectrum. These problems will form part of our future work.

\section*{Acknowledgments}
This work was supported by JST CREST in Japan (Grant No. JPMJCR1661). S.O. was supported by a Grant-in-Aid for Young Scientists from JSPS (Grant No.JP23K13044). A.T. was supported by a Grant-in-Aid for Scientific Research from JSPS in Japan (Grant No. JP23K03281). H.O. was supported by a Grant-in-Aid for Scientific Research from the Japan Society for the Promotion of Science (JSPS) (Grant No. JP21H04988).


\begin{thebibliography}{99}
\bibitem{2DMIdoping1} E. Dagotto, Rev. Mod. Phys. {\bf 766}, 763 (1994).
\bibitem{2DMIdoping2} M. Imada, A. Fujimori, and Y. Tokura, Rev. Mod. Phys. {\bf 770}, 1039 (1998).
\bibitem{2DMIdoping3} A. Damascelli, Z. Hussin, and Z.-X. Shen, Rev. Mod. Phys. {\bf 75}, 473 (2003).
\bibitem{SCSWF} M. Ogata and H. Shiba, Phys. Rev. B {\bf  41}, 2326 (1990).
\bibitem{SCSPL1} A. Parola and S. Sorella, Phys. Rev. Lett. {\bf 64}, 1831 (1990). 
\bibitem{SCSPL2} M. Ogata, T. Sugiyama, and H. Shiba, Phys. Rev. B {\bf  43}, 8401 (1991).
\bibitem{SCCinDoped2DMI1}T. M. Rice and F. C. Zhang, Phys. Rev. B {\bf 39}, 815 (1989).
\bibitem{SCCinDoped2DMI2}  W. Stephan and P. Horsch, Phys. Rev. B {\bf 42}, 8736 (1990).
\bibitem{SCCinDoped2DMI3} J. Wagner, W. Hanke, and D. J. Scalapino, Phys. Rev. B {\bf 43}, 10517 (1991).
\bibitem{SCCinDoped2DMI4} C. -X. Chen and H. -B. Schuttler, Phys. Rev. B {\bf 43}, 3771 (1991).
\bibitem{SCCinDoped2DMI5} E. Dagotto, A. Moreo, F. Ortolani, D. Poilblanc, and J. Riera, Phys. Rev. B {\bf 45}, 10741 (1992).
\bibitem{SCCinDoped2DMI6} R. Strack and D. Vollhardt, Phys. Rev. B {\bf 46}, 13852 (1992).
\bibitem{SCCinDoped2DMI7} W. Metzner, P. Schmit, and D. Vollhardt, Phys. Rev. B {\bf 45}, 2237 (1992).
\bibitem{SCCinDoped2DMI8} D. N. Sheng, Y. C. Chen, and Z. Y. Weng, Phys. Rev. Lett. {\bf 77}, 5102 (1996).
\bibitem{SCCinDoped2DMI9} T. Tohyama and S. Maekawa, Phys. Rev. B {\bf 64}, 212505 (2001).
\bibitem{SCCinDoped2DMI10} G. Sangiovanni, A. Toschi, E. Koch, K. Held, M. Capone, C. Castellani, O. Gunnarsson, S.-K. Mo, J. W. Allen, H.-D. Kim, A. Sekiyama, A. Yamasaki, S. Suga, and P. Metcalf, Phys. Rev. B {\bf 73}, 205121 (2006).
\bibitem{SCCinDoped2DMI11} Z. Zhu, H.-C. Jiang, Y. Qi, C. Tian, and Z.-Y. Weng, Sci. Rep. {\bf 3}, 2586 (2013).
\bibitem{SCCinDoped2DMI12} E. Iyoda and S. Ishihara, Phys. Rev. B {\bf 89}, 125126 (2014).
\bibitem{SCCinDoped2DMI13} T. Terashige, T. Ono, T. Miyamoto, T. Morimoto, H. Yamakawa, N. Kida, T. Ito, T. Sasagawa, T. Tohyama, H. Okamoto, Science Advances {\bf 5}, eaav2187 (2019).
\bibitem{effM1} A. Takahashi, S. Yoshikawa, and M. Aihara, Phys. Rev. B {\bf 65}, 085103 (2002).
\bibitem{effM2} H. Itoh, A. Takahashi, and M. Aihara, Phys. Rev. B {\bf 73}, 075110 (2006).
\bibitem{effM3} A. B. Harris and R. V. Lange, Phys. Rev. 157, 295 (1967).
\bibitem{effM4} A. H. MacDonald, S. M. Girvin, and D. Yoshioka, Phys. Rev. B {\bf 37}, 9753 (1988).
\bibitem{effM5} A. L. Chernyshev, D. Galanakis, P. Phillips, A. V. Rozhkov, and A. M. S. Tremblay, Phys. Rev. B {\bf 70}, 235111 (2004).
\bibitem{Abs1DMI1} E. Jeckelmann, F. Gebhard, and F. H. L. Essler, Phys. Rev. Lett. {\bf 85}, 3910 (2000).
\bibitem{Abs1DMI2} H. Kishida, H. Matsuzaki, H. Okamoto, T. Manabe, M. Yamashita, Y. Taguchi, and Y. Tokura, Nature (London) {\bf 405}, 929 (2000).
\bibitem{Abs1DMI3} Y. Mizuno, K. Tsutsui, T. Tohyama, and S. Maekawa, Phys. Rev. B {\bf 62}, R4769 (2000).
\bibitem{Abs1DMI4} H. Kishida, M. Ono, K. Miura, H. Okamoto, M. Izumi, T. Manako, M. Kawasaki, Y. Taguchi, Y. Tokura, T. Tohyama, K. Tsutsui, and S. Maekawa, Phys. Rev. Lett. {\bf 87}, 177401 (2001).
\bibitem{Abs1DMI5} F. H. L. Essler, F. Gebhard, and E. Jeckelmann, Phys. Rev. B {\bf 64}, 125119 (2001).
\bibitem{Abs1DMI6} A. Takahashi, H. Gomi, and M. Aihara, Phys. Rev. Lett. {\bf 89}, 206402 (2002).
\bibitem{Abs1DMI7} M. Ono, K. Miura, A. Maeda, H. Matsuzaki, H. Kishida, Y. Taguchi, Y. Tokura, M. Yamashita, and H. Okamoto, Phys. Rev. B {\bf 70}, 085101 (2004).
\bibitem{Abs1DMI7.5} N. Maeshima and K. Yonemitsu, J. Phys. Soc. Jpn. {\bf 74}, 2671 (2005).
\bibitem{PIM1D} H. Okamoto, H. Matsuzaki, T. Wakabayashi, Y. Takahashi, and T. Hasegawa, Phys. Rev. Lett. {\bf 98}, 037401 (2007).
\bibitem{Abs1DMI8} A. Takahashi, H. Itoh, and M. Aihara, Phys. Rev. B 77, 205105 (2008).
\bibitem{Srelax1} A. Takahashi, H. Gomi, and M. Aihara, Phys. Rev. Lett. {\bf 89}, 206402 (2002).
\bibitem{Srelax2} A. Takahashi, H. Gomi, and M. Aihara, Phys. Rev. B {\bf 69}, 075116 (2004).
\bibitem{Srelax3} H. Okamoto, T. Miyagoe, K. Kobayashi, H. Uemura, H. Nishioka, H. Matsuzaki, A. Sawa, and Y. Tokura, Phys. Rev. B {\bf 82}, 060513(R) (2010).
\bibitem{Srelax4}  H. Okamoto, T. Miyagoe, K. Kobayashi, H. Uemura, H. Nishioka, H. Matsuzaki, A. Sawa, and Y. Tokura, Phys. Rev. B {\bf 83}, 125102 (2011).
\bibitem{Srelax5} M. Eckstein and P. Werner, Phys. Rev. B {\bf 84}, 035122 (2011).
\bibitem{Srelax6} Z. Lenar\v{c}i\v{c} and P. Prelov\v{s}ek, Phys. Rev. Lett. {\bf 111}, 016401 (2013).
\bibitem{Srelax7} Z. Lenar\v{c}i\v{c} and P. Prelov\v{s}ek, Phys. Rev. B {\bf 90}, 235136 (2014).
\bibitem{Srelax8} D. Gole\v{z}, J. Bon\v{c}a, M. Mierzejewski, and L. Vidmar, Phys. Rev. B {\bf 89}, 165118 (2014).
\bibitem{Srelax9} M. Eckstein and P. Werner, Sci. Rep. {\bf 6}, 21235 (2016).
\bibitem{Srelax10} T. Miyamoto, Y. Matsui, T. Terashige, T. Morimoto, N. Sono, H. Yada, S. Ishihara, Y. Watanabe, S. Adachi, T. Ito, K. Oka, A. Sawa, and H. Okamoto, Nature Communications {\bf 9}, 3948 (2018).
\bibitem{CM} S. Ohmura, A. Takahashi, K. Iwano, T. Yamaguchi, K. Shinjo , T. Tohyama, S. Sota, and H. Okamoto, Phys. Rev. B {\bf 100}, 235134 (2019).
\bibitem{2Dabs1} E. Dagotto, A. Moreo, F. Ortolani, J. Riera, and D. J. Scalapino, Phys. Rev. B {\bf 45}, 10107 (1992).
\bibitem{2Dabs2} P. Wr\'{o}bel and R. Eder, Phys. Rev. B {\bf 66}, 035111 (2002).
\bibitem{2Dabs3} T. Tohyama, Y. Inoue, K. Tsutsui, and S. Maekawa, Phys. Rev. B {\bf 72}, 045113 (2005).
\bibitem{2Dabs4} H. N kano, Y. Takahashi, and M. Imada, J. Phys. Soc. Jpn. {\bf 76}, 034705 (2007).
\bibitem{2Dabs5} C. Taranto, G. Sangiovanni, K. Held, M. Capone, A. Georges, and A. Toschi, Phys. Rev. B {\bf 85}, 085124 (2012).\bibitem{2Dabs6} L. W. Cheuk, M. A. Nichols, K. R. Lawrence, M. Okan, H. Zhang, E. Khatami, N. Trivedi, T. Paiva, M. Rigol, and M. W. Zwierlein, Science {\bf 353}, 1260 (2016).
\bibitem{2Dabs7} E. W. Huang, R. Sheppard, B. Moritz, and T. P. Devereaux, Science {\bf 366}, 987 (2019).
\bibitem{2Dabs8} X.-J. Han, Y. Liu, Z.-Y. Liu, X. Li, J. Chen, H.-J. Liao, Z.-Y. Xie, B. Normand, and T. Xiang, New J. Phys. {\bf 18}, 103004 (2016).
\bibitem{TDMRG} K. Shinjo, Y. Tamaki ,S. Sota, and T. Tohyama, Phys. Rev. B {\bf 104}, 205123 (2021)
\bibitem{2Dabs9} T.-S. Huang , C. L. Baldwin, M. Hafezi, and V. Galitski, Phys. Rev. B {\bf 107}, 075111 (2023).
\bibitem{2Dabs10} O. Mehio, X. Li, H. Ning, Z. Lenar\v{c}i\v{c}, Y. Han, M. Buchhold, Z. Porter, N. J. Laurita, S. D. Wilson, and D. Hsieh, Nature Physics {\bf 19}, 1876 (2023).
\bibitem{IO} K. Iwano and H. Okamoto, Phys. Rev. B {\bf 106}, 075128 (2022)
\bibitem{1D2Dexciton1} W. Stephan and K. Penc, Phys. Rev. B {\bf 54}, R17269(R) (1996).
\bibitem{1D2Dexciton2} F. B. Gallagher and S. Mazumdar, Phys. Rev. B {\bf 56}, 15025 (1997).
\bibitem{1D2Dexciton3} M. Ono, H. Kishida, and H. Okamoto, Phys. Rev. Lett. {\bf 95}, 087401 (2005).
\bibitem{1D2Dexciton4} A. G\"{o}ssling, R. Schmitz, H. Roth, M. W. Haverkort, T. Lorenz, J. A. Mydosh, E. M\"{u}ller-Hartmann, and M. Gr\"{u}ninger, Phys. Rev. B {\bf 78}, 075122 (2008).
\bibitem{1D2Dexciton5} K. A. Al-Hassanieh, F. A. Reboredo, A. E. Feiguin, I. Gonz\'{a}lez, and E. Dagotto, Phys. Rev. Lett. {\bf 100}, 166403 (2008).
\bibitem{1D2Dexciton6} F. Novelli, D. Fausti, J. Reul, F. Cilento, P. H. M. van Loosdrecht, A. A. Nugroho, T. T. M. Palstra, M. Gr\"{u}ninger, and F. Parmigiani, Phys. Rev. B {\bf 86}, 165135 (2012).
\bibitem{1D2Dexciton7} H. Gretarsson, J. P. Clancy, X. Liu, J. P. Hill, E. Bozin, Y. Singh, S. Manni, P. Gegenwart, J. Kim, A. H. Said, D. Casa, T. Gog, M. H. Upton, H.-S. Kim, J. Yu, V. M. Katukuri, L. Hozoi, J. van den Brink, and Y.-J. Kim, Phys. Rev. Lett. {\bf 110}, 076402 (2013).
\bibitem{RSVD} N. Halko, P. G. Martinsson, and J. A. Tropp, SIAM Rev. {\bf 53 (2)}, 217 (2011).
\bibitem{2DabsE1} S. Uchida, T. Ido, H. Takagi, T. Arima, Y. Tokura, and S. Tajima, Phys. Rev. B {\bf 43}, 7942 (1991).
\bibitem{2DabsE2} A. V. Chubukov and D. M. Frenkel, Phys. Rev. B {\bf 52}, 9760 (1995).
\bibitem{1H0D} K. Kusakabe and H. Aoki, Phys. Rev. B {\bf 50}, 12991 (1994).
\bibitem{NagaokaF} Y. Nagaoka, Phys. Rev. {\bf 147}, 392 (1966).
\bibitem{DOS1} J. E. Hirsch, Phys. Rev. B {\bf 31}, 4403 (1985).
\bibitem{DOS2} S. Schmitt, Phys. Rev. B {\bf 82}, 155126 (2010).
\end{thebibliography}
\end{document}